\documentclass[onecolumn,superscriptaddress,nofootinbib,notitlepage]{revtex4-1}
\pdfoutput=1

\usepackage{graphicx}
\usepackage{amsmath}
\usepackage{amssymb}
\usepackage{amsfonts}
\usepackage{dcolumn}
\usepackage{bm}
\usepackage[linktoc=none]{hyperref}
\usepackage[dvipsnames]{xcolor}
\usepackage[utf8]{inputenc}
\usepackage{subfigure}

\usepackage[T1]{fontenc}
\usepackage{pstricks}
\usepackage{color}
\usepackage{multirow}
\usepackage{slashed}
\usepackage{mathtools}

\usepackage{braket}
\usepackage{url}
\usepackage{relsize}
\usepackage{fullpage}
\usepackage{makecell}
\usepackage{blkarray}
\usepackage{fullpage}

\usepackage{feynmp}
\usepackage{feynmp-auto}
\newcommand{\Hi}{\mathcal{H}_i}

\hypersetup{colorlinks,linkcolor={blue},citecolor={teal},urlcolor={violet}}

\newcommand{\rev}[1]{{#1}}

\newcommand{\GRAPPA}{%
Gravitation Astroparticle Physics Amsterdam (GRAPPA),\\
Institute for Theoretical Physics Amsterdam
and Delta Institute for Theoretical Physics,\\
University of Amsterdam, Science Park 904, 1098 XH Amsterdam, The Netherlands}

\newcommand{\SOTON}{Department of Physics and Astronomy, University of Southampton, SO17 1BJ Southampton, United Kingdom}

\begin{document}

\title{Impact of Higgs portal on gravity-mediated production of superheavy dark matter}

\author{Marco Chianese}
\email{m.chianese@uva.nl}
\affiliation{\GRAPPA}

\author{Bowen Fu}
\email{B.Fu@soton.ac.uk}
\affiliation{\SOTON}

\author{Stephen F. King}
\email{king@soton.ac.uk}
\affiliation{\SOTON}

\date{\today}

\begin{abstract}
In the so-called Planckian Interacting Dark Matter (PIDM) scenario, superheavy dark matter particles are produced after inflation by gravity-mediated interactions through the freeze-in mechanism. In the minimal PIDM model, the absence of any additional direct coupling with Standard Model particles is assumed. However, for scalar dark matter particles there is no symmetry that suppresses the Higgs portal coupling. In this paper, we therefore study the impact of a non-zero interaction with the Higgs field on the PIDM paradigm for scalar dark matter. In particular, we fully explore the model parameter space in order to identify the allowed regions where the correct dark matter abundance is achieved. Moreover, we provide the threshold value for the Higgs portal coupling below which the corresponding production processes are sub-dominant and the minimal PIDM scenario is preserved. For a benchmark scalar dark matter mass of $10^{15}$~GeV, we find that the Higgs portal coupling has to be smaller than $5.1 \times 10^{-8}$ ($1.1 \times 10^{-7}$) for instantaneous (non-instantaneous) reheating.
\end{abstract}

\maketitle

\tableofcontents

\section{Introduction \label{sec:intro}}

After almost one century from their first evidence, Dark Matter (DM) particles have revealed themselves only through their gravitational imprints in astrophysical and cosmological systems~\cite{Bertone:2004pz}. On the other hand, no effect due to non-gravitational interactions with known matter has been observed so far in lab-based experiments and indirectly through astrophysical observations~\cite{Undagoitia:2015gya,Kahlhoefer:2017dnp}. The huge experimental effort devoted to DM searches has provided very strong constraints on the parameter space of a variety of potential DM candidates~\cite{Arcadi:2017kky,Arguelles:2019ouk}. Among them, the natural paradigm of Weakly Interacting Massive Particles (WIMPs) thermally produced in the early universe is facing a crisis, thus motivating the investigation of non-WIMP DM candidates~\cite{Bertone:2018xtm}. In this framework, very heavy dark matter candidates with a mass much larger than the electroweak scale, generally known as WIMPzillas~\cite{Chung:1998ua, Kolb:1998ki}, have increasingly gained interest. Differently from WIMPs, these particles cannot be thermally produced through the standard freeze-out mechanism since it would lead to a DM overproduction due to the unitarity limits on the cross-section~\cite{Griest:1989wd}. Hence, heavy DM particles are in general required to be non-thermal relics and to have a very weak interaction with known matter. In this case, one viable DM production mechanism is that DM particles are gradually produced by freeze-in processes~\cite{McDonald:2001vt,Hall:2009bx} (see Ref.~\cite{Bernal:2017kxu} for a review). These DM candidates are generally known as Feebly Interacting Massive particles (FIMPs). Interestingly, it has been shown that the FIMP paradigm allows for a direct link among neutrino physics, leptogenesis and heavy DM particles (FIMPzillas) with a mass at the type-I seesaw scale through the right-handed neutrino portal~\cite{Chianese:2018dsz,Chianese:2019epo}. An alternative scenario is that superheavy DM particles have no direct coupling with SM particles and are instead produced through gravitational interaction only.

Different ways to gravitationally produce superheavy DM particles have been investigated so far. 
First of all, a coupling with gravity makes the fields to develop an effective time-dependent dispersion relation that triggers particle production due to the non-adiabatic expansion of the background spacetime at the end of inflation~\cite{Ford:1986sy,Yajnik:1990un,Chung:1998zb,Kuzmin:1998kk,Chung:1998bt,Peebles:1999fz,Chung:2001cb,Enqvist:2014zqa,Graham:2015rva,Markkanen:2015xuw,Kainulainen:2016vzv,Bertolami:2016ywc,Heikinheimo:2016yds,Ema:2018ucl,Graham:2018jyp,Alonso-Alvarez:2018tus,Fairbairn:2018bsw,Markkanen:2018gcw,Tenkanen:2019aij}. While this production mechanism in general depends on the model of inflation, it is expected to be relevant for DM masses of the order of the Hubble rate $\mathcal{H}_i$ at the end of the inflationary stage. For DM masses larger than $\mathcal{H}_i$, such a DM production is exponentially suppressed~\cite{Chung:1998bt}. Moreover, if the DM is a scalar particle (the main focus of this paper), DM masses smaller than $\mathcal{H}_i$ are excluded by the CMB constraints on isocurvature perturbations~\cite{Chung:2004nh,Nurmi:2015ema,Akrami:2018odb}, unless DM particles are strongly self-interacting~\cite{Markkanen:2018gcw,Tenkanen:2019aij}.

Very recently, it has been pointed out that the FIMP paradigm can be achieved through gravity-mediated processes for DM masses larger than $\mathcal{O}(10^{10})~\mathrm{GeV}$ in the so-called Planckian Interacting Dark Matter (PIDM) scenario~\cite{Garny:2015sjg,Tang:2016vch,Tang:2017hvq,Garny:2017kha,Bernal:2018qlk,Garny:2018grs,Hashiba:2018tbu}. In these effective models, the massless spin-2 graviton couples the stress-energy tensors of SM and DM particles.\footnote{The effect of massive spin-2 mediators has been also investigated in the framework of WIMPs~\cite{Lee:2013bua,Lee:2014caa,Kraml:2017atm,Carrillo-Monteverde:2018phy,Kang:2020huh,Kang:2020yul,Kang:2020afi}.} The only free parameter is the DM mass while the coupling strength is fixed by the equivalence principle. Differently from the gravitational production previously mentioned, the gravity-mediated production occurs after the inflationary stage, and depends on the reheating dynamics. In particular, it depends on the reheating temperature $T_\mathrm{RH}$ of the universe and on the relation between $T_\mathrm{RH}$ and $\mathcal{H}_i$ setting the duration of reheating. In the \textit{minimal} PIDM scenario~\cite{Garny:2015sjg,Garny:2017kha}, it is therefore assumed the absence of any additional non-gravitational interaction of DM particles. However, scalar DM particles inevitably couple to the Higgs field as well since no symmetry exists which forbids the so-called Higgs portal~\cite{Silveira:1985rk,McDonald:1993ex,Burgess:2000yq,Patt:2006fw,Cline:2013gha,Athron:2017kgt,Chanda:2019xyl}. It has already been shown that the DM production through the Higgs portal can easily dominate over the gravitational one from vacuum fluctuations during an inflation epoch driven by a quadratic potential~\cite{Kolb:2017jvz}. However, a similar study regarding the gravity-mediated production after inflation is still missing.

In this paper, we consider the dark matter to be a scalar field focusing on masses larger than $10^9$~GeV and investigate in detail the interplay between the two competitive production processes induced by the gravity-mediated interaction and the Higgs portal in the freeze-in limit. The main aim of the paper is to identify the regions of the model parameter space where one of the two contributions prevails over the other. In particular, we estimate the threshold value for the Higgs portal coupling as a function of the DM mass, below which the corresponding contribution to DM production is negligible and the minimal PIDM scenario is preserved. Moreover, we investigate different reheating scenarios in addition to the instantaneous case.

The paper is organized as follows. In Section~2, we describe the Lagrangian of the model for scalar dark matter. Then, in Section~3 we discuss the dark matter production for instantaneous and non-instantaneous reheating after inflation, and in Section~4 we report our results. Finally, in Section~5 we draw our conclusions.

%%%%%%%%%%%%%%%%%%%%%
\section{The model \label{sec:model}}

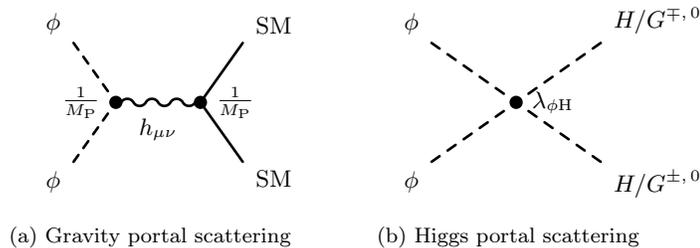
\begin{figure}[t!]
\begin{center}
\subfigure[~Gravity portal scattering]{
\begin{fmffile}{G1}
\fmfframe(20,20)(20,20){
\begin{fmfgraph*}(80,45)
\fmflabel{$\phi$}{i1}
\fmflabel{$\phi$}{i2}
\fmflabel{$\rm SM$}{o2}
\fmflabel{$\rm SM$}{o1}
\fmfv{label=$\tfrac{1}{M_\mathrm{P}}$}{v1}
\fmfv{label=$\tfrac{1}{M_\mathrm{P}}$}{v2}
\fmfleft{i1,i2}
\fmfright{o1,o2}
\fmf{dashes}{i2,v1}
\fmf{dashes}{i1,v1}
\fmf{plain}{v2,o1}
\fmf{plain}{v2,o2}
\fmf{photon,label=$h_{\mu\nu}$}{v1,v2}
\fmfdotn{v}{2}
\end{fmfgraph*}}
\end{fmffile}}
\subfigure[~Higgs portal scattering]{
\begin{fmffile}{H1}
\fmfframe(20,20)(20,20){
\begin{fmfgraph*}(80,45)
\fmflabel{$\phi$}{i1}
\fmflabel{$\phi$}{i2}
\fmflabel{$H/G^{\mp,\,0}$}{o2}
\fmflabel{$H/G^{\pm,\,0}$}{o1}
\fmfv{label=$\lambda_{\rm \phi H}$}{v1}
\fmfleft{i1,i2}
\fmfright{o1,o2}
\fmf{dashes}{i2,v1}
\fmf{dashes}{i1,v1}
\fmf{dashes}{o1,v1}
\fmf{dashes}{o2,v1}
\fmfdotn{v}{1}
\end{fmfgraph*}}
\end{fmffile}}
\end{center}
\caption{\label{fig:Feyn} Scattering processes responsible for scalar dark matter production analyzed in this paper.}
\end{figure}

The minimal model discussed here includes a singlet complex scalar dark matter $\phi$, taken to be odd under a $Z_2$ symmetry to assure its stability. The Lagrangian can be written as 
\begin{equation}
\mathcal{L} = \mathcal{L}_{\rm SM} + \mathcal{L}_{\rm DM} + \mathcal{L}_{\rm EH} + \mathcal{L}_{\rm Gravity~portal} + \mathcal{L}_{\rm Higgs~portal}\,,
\label{eq:lag}
\end{equation}
where $\mathcal{L}_{\rm SM}$ is the SM Lagrangian, $\mathcal{L}_{\rm DM}$ is the free scalar dark matter Lagrangian, and $\mathcal{L}_{\rm EH}$ is the Einstein-Hilbert Lagrangian. The second last term is the coupling the energy-momentum tensors of all the particles through the graviton~\cite{Garny:2017kha} 
\begin{equation}
\mathcal{L}_{\rm Gravity~portal} = \frac{\sqrt{8\pi}}{2\,M_\mathrm{P}} h^{\mu\nu}(T^{\rm SM}_{\mu\nu} + T^{\rm DM}_{\mu\nu})\,,
\label{eq:laggr}
\end{equation}
where $M_\mathrm{P} = 1.2 \times 10^{19}~\mathrm{GeV}$ is the non-reduced Planck mass. Finally, the last term in Eq.~\eqref{eq:laggr} is the coupling between the SM Higgs and the dark scalar
\begin{equation}
\mathcal{L}_{\rm Higgs~portal} = \lambda_{\rm  \phi H}  \left| \phi \right|^2  \left| H \right|^2\,,
\label{eq:lagh}
\end{equation}
where $\lambda_{\rm  \phi H}$ is the Higgs portal coupling. In this model, a non-minimal gravitational coupling with the Ricci curvature is allowed~\cite{Chung:1998zb,Kuzmin:1998kk,Garny:2015sjg,Garny:2017kha}. As discussed in the Introduction, such a coupling would induce the gravitational production from vacuum fluctuations which depends on the inflationary model and is exponentially suppressed for $m_\phi > \mathcal{H}_i$, where the gravity-mediated processes are dominant~\cite{Garny:2015sjg,Garny:2017kha}. For this reason, we do not include it in the present work. Hence, the dark matter can interact with the Standard Model particles through both the gravity and the Higgs portals. The Feynman diagram for the scattering processes contributing to dark matter production is shown in Fig.~\ref{fig:Feyn}.
The general formula for the amplitude $\mathcal{M}$ in terms of the stress-energy tensors of DM and SM particles is 
\begin{equation}
\mathcal{M}_\mathrm{Gravity} =  \frac{-i\,8\pi}{s \, M_\mathrm{P}^2} \left(T^{\mu\nu}_\mathrm{SM}T_{\mu\nu}^\mathrm{DM}-\frac{1}{2}T_\mathrm{SM} T_\mathrm{DM}\right)\,,
\end{equation}
where $T^{\mu\nu}_\mathrm{SM}$ and $T_{\mu\nu}^\mathrm{DM}$ are the stress-energy tensors of SM and DM particles, respectively, and $T_\mathrm{SM}$ and $T_\mathrm{DM}$ are their traces. The quantity $T^{\mu\nu}_\mathrm{DM}$ takes the expression
\begin{equation}
T^{\mu\nu}_\mathrm{DM} = \frac12 \left[k_1^\mu k_2^\nu + k_1^\nu k_2^\mu - \eta^{\mu\nu}(k_1\cdot k_2 + m_\phi^2)\right]\,,
\end{equation}
while the expressions for $T^{\mu\nu}_\mathrm{SM}$ depends on the Lorentz nature of the standard model particles, so having
\begin{eqnarray}
T^{\mu\nu}_\mathrm{SM}  =  
\begin{dcases} 
\frac{1}{2} (p_1^\mu p_2^\nu + p_1^\nu p_2^\mu - \eta^{\mu\nu}p_1\cdot p_2) 
& \text{for scalar}\\
\frac{1}{4} \overline{u}(p_2)\left[\gamma^\mu (p_1^\nu - p_2^\nu) + \gamma^\nu (p_1^\mu - p_2^\mu)\right] u(p_1) 
& \text{for spinor} \\
\frac{1}{2} \left[ \epsilon_1\cdot\epsilon_2(p_1^\mu p_2^\nu + p_1^\nu p_2^\mu) 
- p_1\cdot\epsilon_2(\epsilon_1^\mu p_2^\nu + \epsilon_1^\nu p_2^\mu) 
- \epsilon_1\cdot p_2(p_1^\mu \epsilon_2^\nu + p_1^\nu \epsilon_2^\mu) \right.&\\
\quad \left. + p_1\cdot p_2(\epsilon_1^\mu \epsilon_2^\nu + \epsilon_1^\nu \epsilon_2^\mu) + \eta^{\mu\nu}(\epsilon_1\cdot p_2\, p_1\cdot\epsilon_2 - p_1\cdot p_2 \epsilon_2\cdot\epsilon_1) \right]  
& \text{for vector}
\end{dcases}
\end{eqnarray}
where $k_i$ and $p_i$ are the 4-momenta of DM and SM particles, respectively. The amplitude for the Higgs portal processes takes instead the following very simple expression
\begin{equation}
\mathcal{M}_\mathrm{Higgs} =  -i \,\lambda_{\rm \phi H}\,.
\end{equation}

%%%%%%%%%%%%%%%%%%
\section{Dark matter production}

We analyze the scalar dark matter production driven by the two processes reported in Fig.~\ref{fig:Feyn} after the inflationary stage. The reheating process can be characterized by defining the quantity  $\gamma \equiv \sqrt{\Gamma/\Hi}$, where $\Gamma$ and $\mathcal{H}_i$ are the decay rate of the inflaton and the Hubble rate at the end of inflation, respectively. Depending on these quantities, it can be either instantaneous ($\gamma = 1$) or non-instantaneous ($\gamma < 1$). The former can be achieved in nonperturbative reheating scenarios~\cite{Felder:1998vq}, while the latter is typically obtained in perturbative reheating models. In the scenario of perturbative reheating, the reheating temperature is given by the condition $\Gamma = \mathcal{H}_{\rm RH}$, which provides a straightforward relation between Hubble parameter at the end of inflation and reheating temperature
\begin{eqnarray}
\Hi  = \gamma^{-2} \sqrt{\frac{4\pi^3{g_*}_{\rm RH}}{45}} \frac{T_{\rm RH}^2}{M_{\rm P}}\,,
\label{eq:HTRH}
\end{eqnarray}
where ${g_*}_{\rm RH}$ is the degree of freedom at the end of reheating. We take ${g_*}_{\rm RH} = 106.75$ according to the SM matter content. The current bound on tensor modes $r<0.07$ at 95\% C.L. deduced by CMB measurements~\cite{Akrami:2018odb, Ade:2015lrj, Ade:2018gkx} provides a constraint on the Hubble rate at the end of the inflation,
\begin{equation}
\mathcal{H}_i \lesssim 6.1 \times 10^{13}~\mathrm{GeV} \,,
\end{equation}
which using Eq.~\eqref{eq:HTRH} translates into the following bound on the reheating temperature
\begin{equation}
 T_\mathrm{RH} \lesssim 6.5 \times 10^{15}\,\gamma~\mathrm{GeV}\,.
 \label{eq:limitTRH}
\end{equation}
Moreover, we assume that SM particles are in local thermodynamic equilibrium. This is indeed likely to happen because the SM particles dominate the thermal bath, unless the inflaton decays into additional particles that are weakly coupled to all SM particles~\cite{Chung:1998rq,Kurkela:2014tea}. On the other hand, the dark scalar is far from thermal equilibrium in all the processes, further justifying the freeze-in regime., and consequently their number densities are described by Maxwell-Boltzmann distributions. For the sake of simplicity, we assume that the number densities of particles in thermal equilibrium are described by Maxwell-Boltzmann distributions. In general, this approximation is not reliable in the relativistic limit where the Fermi-Dirac and Bose-Einstein statistics should be instead taken into account. For example, in case of scalar dark matter, it has recently been shown that  in the limit $m_\phi \ll T$ the correct production rates related to the Higgs portal are up to a factor of 1.5 larger than the ones computed in the naive Maxwell-Boltzmann approximation~\cite{Lebedev:2019ton}. However, as shown in Ref.~\cite{Lebedev:2019ton}, such an enhancement is approximately compensated by other effects such as thermal mass corrections. This further justifies our simplistic approach in the current work. A more detailed calculation is left for future work.

In the next subsections, we will describe the Boltzmann equations relevant to the dark matter production in the instantaneous and non-instantaneous reheating scenarios. The solution of these Boltzmann equations provides the total DM relic abundance as
\begin{equation}
\Omega_{\rm DM}h^2 =  \frac{2 \, \mathfrak{s}_0 \, m_\phi  \,Y_{\phi,0}}{\rho_{\rm crit}/h^2}  \,,
\label{eq:omega}
\end{equation}
where $\mathfrak{s}_0=2891.2\,{\rm cm^3}$ and $\rho_{\rm crit}/h^2 = 1.054 \times10^{-5} {\rm GeV \, cm^{-3}}$ are respectively the today’s entropy density and the critical density~\cite{Tanabashi:2018oca}, $Y_{\phi,0}$ is the today's DM yield (ratio of number density and entropy density), and the factor of 2 takes into account the contribution of DM anti-particles. This predicted value is required to match the experimental value $\Omega_{\rm DM}h^2 = 0.120 \pm 0.001$ at 68\% C.L.~\cite{Aghanim:2018eyx}.

%%%%%%%%%%%%%%%%%%%%%%%%%%%
\subsection{Instantaneous reheating}

In the case of instantaneous reheating ($\gamma=1$), in the freeze-in limit the Boltzmann equation describing the evolution of dark matter yield after reheating can be reduced to 
\begin{eqnarray}
\mathcal{H}\,T\left(1+\frac{T}{3 g^\mathfrak{s}_*}\frac{\mathrm{d} g^\mathfrak{s}_*}{\mathrm{d} T}\right)^{-1} \frac{\mathrm{d} Y_\phi}{\mathrm{d} T} & = & 
- \mathfrak{s} \left<\sigma\, v\right>_{\phi\phi} \left({Y_\phi^{\rm eq}}\right)^2  \,, \label{eq:Boltzmann} 
\end{eqnarray}
where $\mathcal{H}$ and $\mathfrak{s}$ are the Hubble parameter and the entropy density, and $g^\mathfrak{s}_*$ is the relativistic degrees of freedom which in general depend on the temperature. However, it has been checked that changes in $g^\mathfrak{s}_*$ only provides a correction of order $10^{-3}$, therefore it is a good approximation to take $g^\mathfrak{s}_*=106.75$ according to the SM particle content. The yield $Y_\phi$ is defined as $n_\phi / \mathfrak{s}$ and $Y_\phi^{\rm eq}$ is its value at thermal equilibrium. Assuming zero DM number density at the end of inflation, the today's yield of DM particles is simply given by
\begin{eqnarray}
Y_{\phi,0} &  = & \int_{0}^{T_{\rm RH}} \mathrm{d} T \, \frac{\mathfrak{s}}{\mathcal{H}\,T} 
\left<\sigma\, v\right>_{\phi\phi} \left({Y_\phi^{\rm eq}}\right)^2\,.
\end{eqnarray}
The total thermally averaged cross-section $\left<\sigma\, v\right>_{\phi\phi}$, computed according to Ref.~\cite{Edsjo:1997bg}, is given by the sum of the contributions from both gravity and Higgs portal processes. Hence, we have
\begin{equation}
\left<\sigma\, v\right>_{\phi\phi} = \left<\sigma\, v\right>_{\rm Gravity} + \left<\sigma\, v\right>_{\rm Higgs}\,.
\label{eq:cross}
\end{equation}
The first term is given by
\begin{equation}
\langle \sigma v \rangle _{\rm Gravity}= 4 \langle \sigma_{0\rightarrow0} \,v \rangle + 45 \langle \sigma_{1/2\rightarrow0}  \,v \rangle + 12 \langle \sigma_{1\rightarrow0} \,v \rangle 
\simeq
\begin{dcases}
    \frac{\pi m_\phi^2}{2 M_\mathrm{P}^4}    & m_\phi \gg T\\
    \frac{916 \pi T^2}{5 M_\mathrm{P}^4}     & m_\phi \ll T
\end{dcases} 
\label{eq:avggr} \,,
\end{equation}
where we sum over the production processes of scalar DM particles from the whole SM matter content with spin 0, $1/2$ and 1. They take the following expressions
\begin{eqnarray}
\langle \sigma_{0\rightarrow0} \rangle & = & \frac{\pi m_\phi^2}{40 M_\mathrm{P}^4}  \left( 3 \frac{K_1^2}{K_2^2} + 2 +4 \frac{T}{m_\phi} \frac{K_1}{K_2} + 8 \frac{T^2}{m_\phi^2} \right)
\simeq 
\begin{dcases}
    \frac{\pi m_\phi^2}{8 M_\mathrm{P}^4}  & m_\phi \gg T \,\\
    \frac{\pi T^2}{5 M_\mathrm{P}^4}   & m_\phi \ll T\,
\end{dcases} \,, \\ 
&& \nonumber \\
\langle \sigma_{1/2\rightarrow0}  \rangle = 
\langle \sigma_{1\rightarrow0} \rangle  & = &  \frac{8\pi m_\phi^2}{15 M_\mathrm{P}^4} \left( \frac{K_1^2}{K_2^2}-1 +3 \frac{T}{m_\phi} \frac{K_1}{K_2} + 6\frac{T^2}{m_\phi^2} \right) 
\simeq
\begin{dcases}
    \frac{4\pi T^2}{ M_\mathrm{P}^4}  & m_\phi  \gg T \,\\
    \frac{16 \pi T^2}{5 M_\mathrm{P}^4}   & m_\phi \ll  T \,
\end{dcases} \,. 
\end{eqnarray}
where $K_1$ and $K_2$ are the first and second modified Bessel functions of the second kind with argument $m_\phi / T$. Similarly, the thermally averaged cross-section for the Higgs portal scatterings is given by
\begin{eqnarray}
\langle \sigma v \rangle_{\rm Higgs} = \frac{ \lambda_{\rm  \phi H}^2}{8 \pi m_\phi^2} \frac{K_1^2}{K_2^2} \simeq
\begin{dcases}
    \frac{ \lambda_{\rm  \phi H}^2}{8 \pi \, m_\phi^2}  & m_\phi \gg T \\
     \frac{ \lambda_{\rm  \phi H}^2}{32 \pi  \,T^2}      & m_\phi \ll T
\end{dcases} \, .\label{eq:AvgHiggs}
\end{eqnarray}
Therefore, for $m_\phi \gg T$ both the cross-sections are independent of temperature. On the other hand, at high temperatures ($m_\phi \ll T$) the gravity-mediated process becomes more efficient as the temperature increases, while the Higgs scattering production becomes less efficient. As a result, the Higgs portal coupling achieving the correct DM relic abundance is expected to be very sensitive to the reheating temperate when $m_\phi \ll T$.

According to Eq.~\eqref{eq:cross}, we can separately estimate the contributions to the DM yield from the two different processes by splitting the Boltzmann equation~\eqref{eq:Boltzmann} into two parts as
\begin{eqnarray}
\left. \frac{\mathrm{d}\, Y_\phi}{\mathrm{d}\, T} \right|_{\rm Gravity} &\simeq &
\begin{dcases}
\frac{T}{\mathcal{H}\,\mathfrak{s}} \frac{m_\phi^6}{8 \pi^3 M_\mathrm{P}^4} K_1^2  \propto \frac{m_\phi^6}{M_\mathrm{P}^3 \, T^4} K_1^2 
& m_\phi \gg T \\
\frac{T}{\mathcal{H}\,\mathfrak{s}} \frac{916 \, m_\phi^2 \, T^4}{5 \pi^3 M_\mathrm{P}^4}   K_1^2  \propto  \frac{m_\phi^2}{M_\mathrm{P}^3} K_1^2 
& m_\phi \ll T 
\end{dcases} \,, \\
&& \nonumber \\
\left. \frac{\mathrm{d}\, Y_\phi}{\mathrm{d}\, T} \right|_{\rm Higgs}  &\simeq & \frac{T}{\mathcal{H}\,\mathfrak{s}}  \frac{\lambda_{\rm  \phi H}^2 \, m_\phi^2}{8 \pi^5}  K_1^2  \propto  \frac{\lambda_{\rm  \phi H}^2 \, m_\phi^2 \, M_\mathrm{P}}{T^4} K_1^2  \,.
\end{eqnarray}
For superheavy scalar dark matter with mass larger than the reheating temperature ($m_\phi \gg T_{\rm RH}$), the contributions to the yield of dark matter are 
\begin{eqnarray}
Y_{\phi, 0}^{\rm Gravity}  \simeq  \int_{0}^{T_{\rm RH}} \mathrm{d}T \frac{T}{\mathcal{H}\,\mathfrak{s}} \frac{m_\phi^6}{8 \pi^3 M_\mathrm{P}^4} K_1^2  \quad\text{and}\quad
Y_{\phi, 0}^{\rm Higgs}  \simeq  \int_{0}^{T_{\rm RH}} \mathrm{d}T  \frac{T}{\mathcal{H}\,\mathfrak{s}}  \frac{\lambda_{\rm  \phi H}^2 m_\phi^2}{8 \pi^5}  K_1^2\,.
\end{eqnarray}
Observing that the dependences on temperature of the two contributions are the same, the ratio
\begin{eqnarray}
\frac{Y_{\phi, 0}^{\rm Gravity} }{Y_{\phi, 0}^{\rm Higgs}} = \frac{ \pi^2}{\lambda_{\rm  \phi H}^2} \frac{ m_\phi^4}{M_\mathrm{P}^4} \qquad {\rm for} \quad m_\phi \gg T_{\rm RH}\,,
\label{eq:rcontri}
\end{eqnarray}
is independent of the reheating temperature: it only depends on the dark matter mass and the Higgs portal coupling. Then it is straightforward to estimate that the gravity production and Higgs portal production switch their dominance at around
\begin{equation}
\lambda^{\rm equality}_{\rm  \phi H} = \pi \left(\frac{\, m_\phi^2}{M_\mathrm{P}^2}\right)  \qquad {\rm for} \quad m_\phi \gg T_{\rm RH} \,,\label{eq:coupling1}
\end{equation}
for which the two different processes equally contribute to DM production. Such a value for the Higgs portal coupling at equality only depends on the dark matter mass. 

On the other hand, for scalar dark matter with mass smaller than the reheating temperature ($m_\phi \ll T_{\rm RH}$), one can show that the ratio of the contributions from the two different scattering processes actually only depends on the Higgs portal coupling. For the gravity-mediated process, when $m_\phi \ll T_{\rm RH}$, the production is dominated by the ultraviolet part, typically at a temperature around the reheating temperature. Such behaviour is indeed in agreement with the so-called ultraviolet freeze-in scenario~\cite{Elahi:2014fsa}. The yield due to gravity production can be estimated as 
\begin{eqnarray}
&Y_{\phi,0}^\text{Gravity} & \simeq  \int_{0.1 T_{\rm RH}}^{T_{\rm RH}} \mathrm{d}T\frac{T}{\mathcal{H}\,\mathfrak{s}} \frac{916 \, T^6}{5 \pi^3 M_\mathrm{P}^4}    \simeq C \times \frac{916}{15  \, \pi^3} \left(\frac{T_{\rm RH}}{M_\mathrm{P}}\right)^3\,,\label{eq:grY2}
\end{eqnarray}
with
\begin{equation}
C=\sqrt{\frac{45}{4\pi^3{g_*}_{\rm RH}}}\frac{45}{2\pi^2{g_*}_{\rm RH}} \simeq 1.25\times10^{-3}\,.
\end{equation}
For the Higgs portal process, the production mainly happens at temperature around the scalar mass, 
\begin{eqnarray}
&Y_{\phi,0}^\text{Higgs} &  \simeq \int_{0.1 m_\phi}^{10 m_\phi} \mathrm{d}T  \frac{T}{\mathcal{H}\,\mathfrak{s}}  \frac{\lambda_{\rm  \phi H}^2 m_\phi^2}{8 \pi^5} K_1^2  \simeq   C \times \frac{\lambda_{\rm  \phi H}^2}{8 \pi^5} \left(\frac{M_\mathrm{P}}{m_\phi}\right) \,.
\end{eqnarray}
Together, the ratio of the contributions is
\begin{eqnarray}
\frac{Y_{\phi, 0}^{\rm Gravity} }{Y_{\phi, 0}^{\rm Higgs}} = \frac{ 7328\, \pi^2}{45\, \lambda_{\rm  \phi H}^2} \left(\frac{ m_\phi\, T_{\rm RH}^3}{M_\mathrm{P}^4}\right) \qquad {\rm for} \quad m_\phi \ll T_{\rm RH}\,.
\label{eq:rcontri2}
\end{eqnarray}
At first sight, the expression above seems to depend on three parameters: the Higgs portal coupling $\lambda_{\rm  \phi H}$, the scalar mass $m_\phi$ and the reheating temperature $T_{\rm RH}$. However, the latter two quantities are related to each other. From the estimated yield for gravity-mediated production in Eq.~\eqref{eq:grY2} and the DM relic abundance given in Eq.~\eqref{eq:omega}, one gets
\begin{eqnarray}
m_\phi \,T_{\rm RH}^3 = \frac{15  \pi^3 M_\mathrm{P}^3}{916 \,C} \frac{\,\Omega_{\rm DM}\rho_{\rm crit}}{2 \, \mathfrak{s}_0} \frac{Y_{\phi, 0}^{\rm Gravity} }{Y_{\phi, 0}^{\rm Gravity} + Y_{\phi, 0}^{\rm Higgs}}\,.
\end{eqnarray}
Then, Eq.~\eqref{eq:rcontri2} turns into 
\begin{eqnarray}
\frac{Y_{\phi, 0}^{\rm Gravity} }{Y_{\phi, 0}^{\rm Higgs}} = \frac{ 4\, \pi^5 \,\Omega_{\rm DM}\rho_{\rm crit}}{\lambda_{\rm  \phi H}^2\, C\, M_\mathrm{P} \, \mathfrak{s}_0} -1\,.
\end{eqnarray}
which only depends on the Higgs portal coupling $\lambda_{\rm  \phi H}$. Hence, for $m_\phi \ll T_{\rm RH}$ the Higgs portal coupling at equality is estimated to be 
\begin{eqnarray}
\lambda^{\rm equality}_{\rm  \phi H} = \sqrt{ \frac{2 \pi^5 \, \Omega_{\rm DM}\rho_{\rm crit}}{C \, M_\mathrm{P} \, \mathfrak{s}_0}} \simeq 4.18\times10^{-12} \qquad {\rm for} \quad m_\phi \ll T_{\rm RH} \,.
\label{eq:coupling2}
\end{eqnarray}
Such a value, along with the one reported in Eq.~\eqref{eq:coupling1} in the opposite limit $m_\phi \gg T_{\rm RH} $, represents the threshold value for the Higgs portal coupling below which the production of DM particles is predominantly driven by gravity-mediated processes and the Planckian Interactive Dark Matter paradigm is preserved. These threshold values will be further investigated by numerical analysis in the next Section.

%%%%%%%%%%%%%%%%%%%%%%%%
\subsection{Non-instantaneous reheating}

Let us now focus on the non-instantaneous reheating case ($\gamma < 1$) and discuss the DM production during reheating. We assume the scenario of perturbative reheating and follow the approach discussed in Ref.~\cite{Giudice:2000ex}. In the case of feeble interactions where the dark matter never dominates the energy density, the effect of inflaton and radiation on the DM production is negligible. However, the photon temperature is no longer a good variable for integration because the Hubble parameter is not a bijective function of temperature during reheating. As a result, an alternative form of Boltzmann equation equivalent to Eq.~\eqref{eq:Boltzmann} is more convenient for calculation
\begin{eqnarray}
\frac{\mathrm{d} X_\phi}{\mathrm{d} a} & = & 
\frac{a^2}{T_{\rm RH}^3 \mathcal{H}} \left<\sigma\, v\right>_{\phi\phi} \left({n_\phi^{\rm eq}}\right)^2   \,, 
\label{eq:phi} 
\end{eqnarray}
where $X_\phi=n_\phi a^3 / T_{\rm RH}^3$ and $a$ is the scalar factor normalised at the end of inflation. During reheating, the Hubble parameter and temperature behave differently
\begin{eqnarray}
\mathcal{H} &=& \Hi  \left(\frac{a}{a_i} \right)^{-3(1+w)/2}\,,\\
T(a) & = & \left(\frac{2}{5+3w}\right)^{1/4} \left(\frac{{g_*}_{\rm RH}}{{g_*}(a)}\right)^{1/4} {\gamma}^{-1/2} \left(\left(\frac{a}{a_i} \right)^{-3(1-w)/2}-\left(\frac{a}{a_i} \right)^{-4}\right)^{1/4} T_{\rm RH}\,,\label{eq:temdur}
\end{eqnarray}
where ${g_*}(a)$ is the degree of freedom as a function of the scalar factor and $w$ parameterised the effective equation of state of the inflaton field dynamics. As can be seen in the above equations, the temperature has a maximum $T_{\rm max}$ during reheating at $a/a_i=\left({8}/{3(1-w)}\right)^{2/(3w+5)}$, after which it decreases as $(a/a_i)^{-3(1-w)/8}$ until the end of reheating. Within the assumption of matter dominance during reheating ($w=0$), the maximum temperature is $T_{\rm max}=0.61 \, \gamma^{-1/2} \, T_\mathrm{RH}$. According to Ref.~\cite{Bernal:2019mhf}, the gravity-mediated production can be strongly enhanced at $T_\mathrm{max}$ in the limit $m_\phi \ll T_\mathrm{max}$ due to the ultraviolet freeze-in behaviour~\cite{Elahi:2014fsa}. On the other hand, if the mass of the dark scalar is much larger than the maximum temperature, the relation in Eq.~\eqref{eq:rcontri} still holds. This means that  the Higgs portal coupling, for which the contributions of the two production processes are equal, is determined only by the scalar mass. The scale factor at the reheating temperature is given by
\begin{eqnarray}
a_{\rm RH} = a_i \gamma^{-\frac{4}{3(1+w)}} \,.
\end{eqnarray}
After reheating, we can then use the Boltzmann equation as reported in Eq.~\eqref{eq:Boltzmann}.

%%%%%%%%%%%%%%%%%%%%%%%%%%%
\subsection{Thermalisation constraint}
\rev{The freeze-in production mechanism requires that dark matter particles never reach the thermal equilibrium with SM particles in the early universe. However, when the Higgs portal coupling is large enough, the interaction could be strong enough to ensure thermal equilibrium between the two sectors, so leading to an over-production of superheavy dark matter particles due to the thermal freeze-out regime. In general, the non-thermalisation condition is conservatively satisfied when the interaction rates are smaller than the expansion rate of the universe. Hence, in the minimal model of scalar dark matter, it reads
\begin{equation}
\frac{n^{\rm eq}_{\phi} \langle \sigma v \rangle_{\rm Higgs}}{\mathcal{H}} <1 \,.
\label{eq:therm}
\end{equation}
Such a requirement therefore provides a constraint on the parameter space. With the thermal averaged cross-section in Eq.~\eqref{eq:AvgHiggs}, in the scenario of instantaneous reheating the ratio on the left-hand side of Eq.~\eqref{eq:therm} can be recast as
\begin{eqnarray}
\frac{n^{\rm eq}_{\phi} \langle \sigma v \rangle_{\rm Higgs}}{\mathcal{H}} =
\sqrt{\frac{45}{4\pi^3{g_*}}}\frac{ \lambda_{\rm  \phi H}^2}{16\pi^3} \frac{M_{\rm P}}{T} \frac{K_1^2\left(m_\phi/T\right)}{K_2\left(m_\phi/T\right)} \simeq
\begin{dcases}
7.2 \times 10^{14}  \, \lambda_{\rm  \phi H}^2 \left(\frac{\mathrm{GeV}}{T}\right)  & m_\phi \ll T \\
1.4 \times 10^{15} \, \lambda_{\rm  \phi H}^2 \left(\frac{\mathrm{GeV}}{T}\right) K_1\left(m_\phi/T\right)  &  m_\phi \gg T 
\end{dcases} \,. \label{eq:thermalcon}
\end{eqnarray}
Depending on the reheating temperature $T_\mathrm{RH}$, which corresponds to the maximum temperature of the universe in the instantaneous reheating scenario, the ratio is maximized at different values of the temperature $T$. In particular, one can show by examining the behaviour of the Bessel functions which depend on the ratio $m_\phi/T$ that, when $m_\phi < T_\mathrm{RH}$, the ratio between the Higgs interaction rate and the Hubble parameter takes its maximum value at $T \simeq m_\phi$.\footnote{The reason is that the quantity in Eq.~\eqref{eq:thermalcon} increases as the temperature $T$ decreases when $m_\phi < T$, while it increases as $T$ increases for $m_\phi > T$, therefore when $m_\phi < T_\mathrm{RH}$ it is maximised at $T \simeq m_\phi$.} Hence, in this case the non-thermalisation condition provides the following upper limit to the Higgs portal coupling
\begin{equation}
\lambda_{\rm  \phi H}^2 \lesssim 3.1 \times 10^{-15}  \,\left(\frac{m_\phi}{\mathrm{GeV}}\right) \qquad \text{for} \quad m_\phi < T_\mathrm{RH}\,.
\end{equation}
On the other hand, when $m_\phi > T_\mathrm{RH}$, the expression in Eq.~\eqref{eq:thermalcon} takes its maximum value at $T=T_\mathrm{RH}$ due to the Boltzmann suppression encoded by the Bessel function $K_1$. In this case, the non-thermalisation condition is satisfied when
\begin{equation}
\lambda_{\rm  \phi H}^2 \lesssim 7.0 \times 10^{-16}  \,\left(\frac{T_\mathrm{RH}}{\mathrm{GeV}}\right) \frac{1}{K_1\left(m_\phi/T_{\rm RH}\right)}  \qquad \text{for} \quad m_\phi > T_\mathrm{RH}\,.
\end{equation}
In the scenario of non-instantaneous reheating, the maximum of the ratio $n^{\rm eq}_{\phi} \langle \sigma v \rangle_{\rm Higgs}/\mathcal{H}$ appears during reheating. In particular, during reheating we have
\begin{equation}
\frac{n^{\rm eq}_{\phi} \langle \sigma v \rangle_{\rm Higgs}}{\mathcal{H}}= 
a^\frac32  \gamma^2 \sqrt{\frac{45}{4\pi^3{g_*}}}\frac{\lambda_{\rm  \phi H}^2}{16\pi^3} \frac{M_{\rm P}\,T(a)}{T_{\rm RH}^2} \frac{K_1^2\left(m_\phi/T(a)\right)}{K_2\left(m_\phi/T(a)\right)} \,,
\label{eq:non-inst_ratio}
\end{equation}
with $T(a)$ given by Eq.~\eqref{eq:temdur}. The maximum of the ratio does not appear at the maximum temperature $T_\mathrm{max}$ but instead it is reached at some particular value of the scale factor $a$ depending on the ratio between the scalar mass and the reheating temperature. If the scalar mass is smaller than the temperature, the ratio~\eqref{eq:non-inst_ratio} is a monotonically increasing function of the scale factor $a$ and its maximum value appears at the reheating temperature. In this case, the constraint on the Higgs portal coupling provided by the non-thermalisation condition is computed numerically as a function of $m_\phi$ and $T_\mathrm{RH}$.
}

%%%%%%%%%%%%%%%%%%%%
%%%%%%%%%%%%%%%%%%%%
\section{Results \label{sec:results}}

In the case of instantaneous reheating ($\gamma = 1$), there are three free parameters: the dark scalar mass $m_\phi$, the reheating temperature $T_\mathrm{RH}$, and the Higgs portal coupling $\lambda_{\rm  \phi H}$. These parameters are constrained by requiring the production of the correct DM relic abundance. For non-instantaneous reheating, there is the additional free parameter $\gamma$, parameterising the ratio of the Hubble parameter at the end of inflation and at the end of reheating. We consider two benchmark cases of $\gamma=0.1$ and $\gamma=0.01$. During reheating, the universe is assumed to be matter-dominated, considering the coherent oscillations of the inflaton field.

\begin{figure}[t!]
\begin{center}
\subfigure[]{\includegraphics[width=0.435\textwidth]{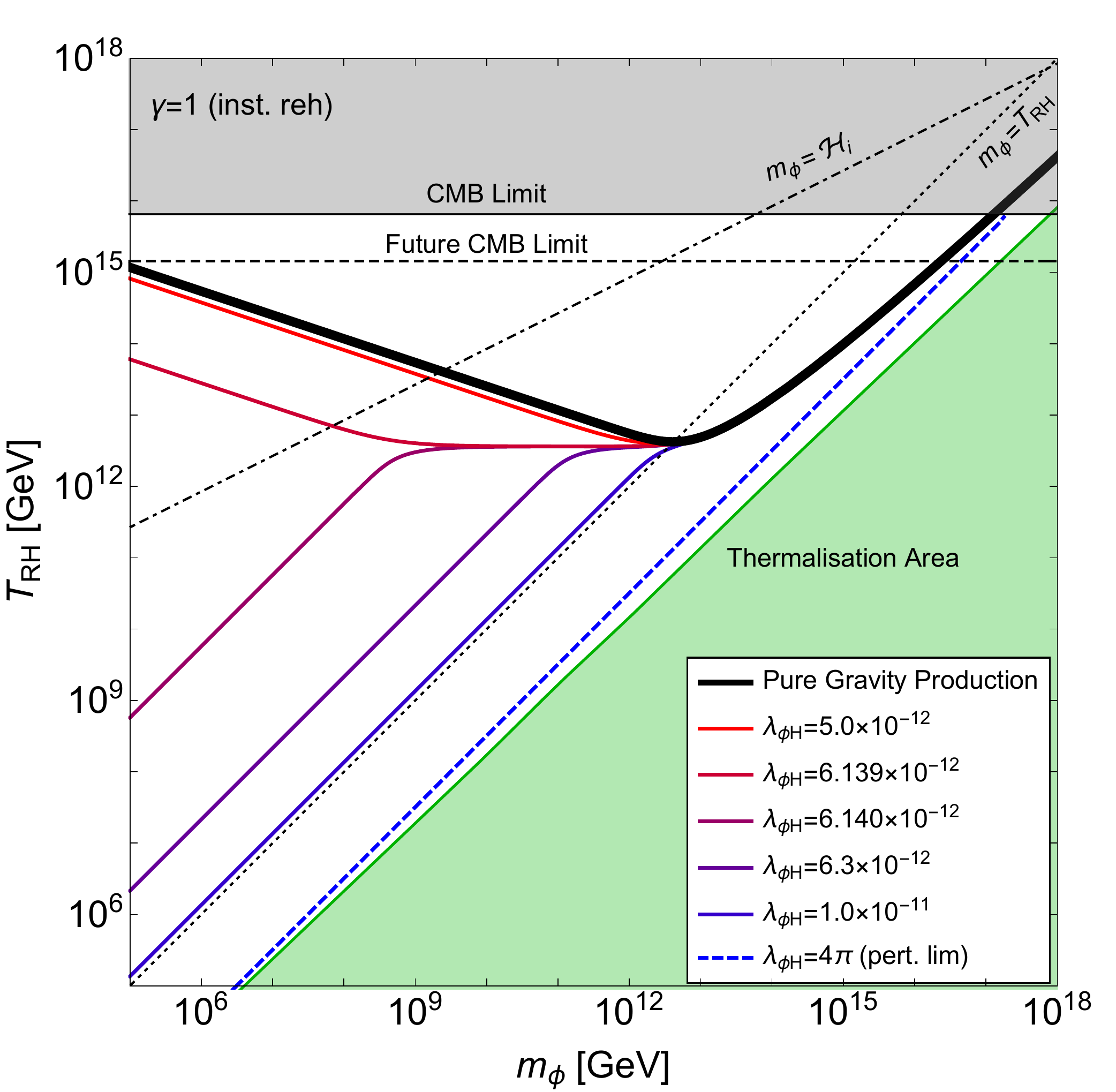}\label{fig:TRHvsM1}}
\subfigure[]{\includegraphics[width=0.55\textwidth]{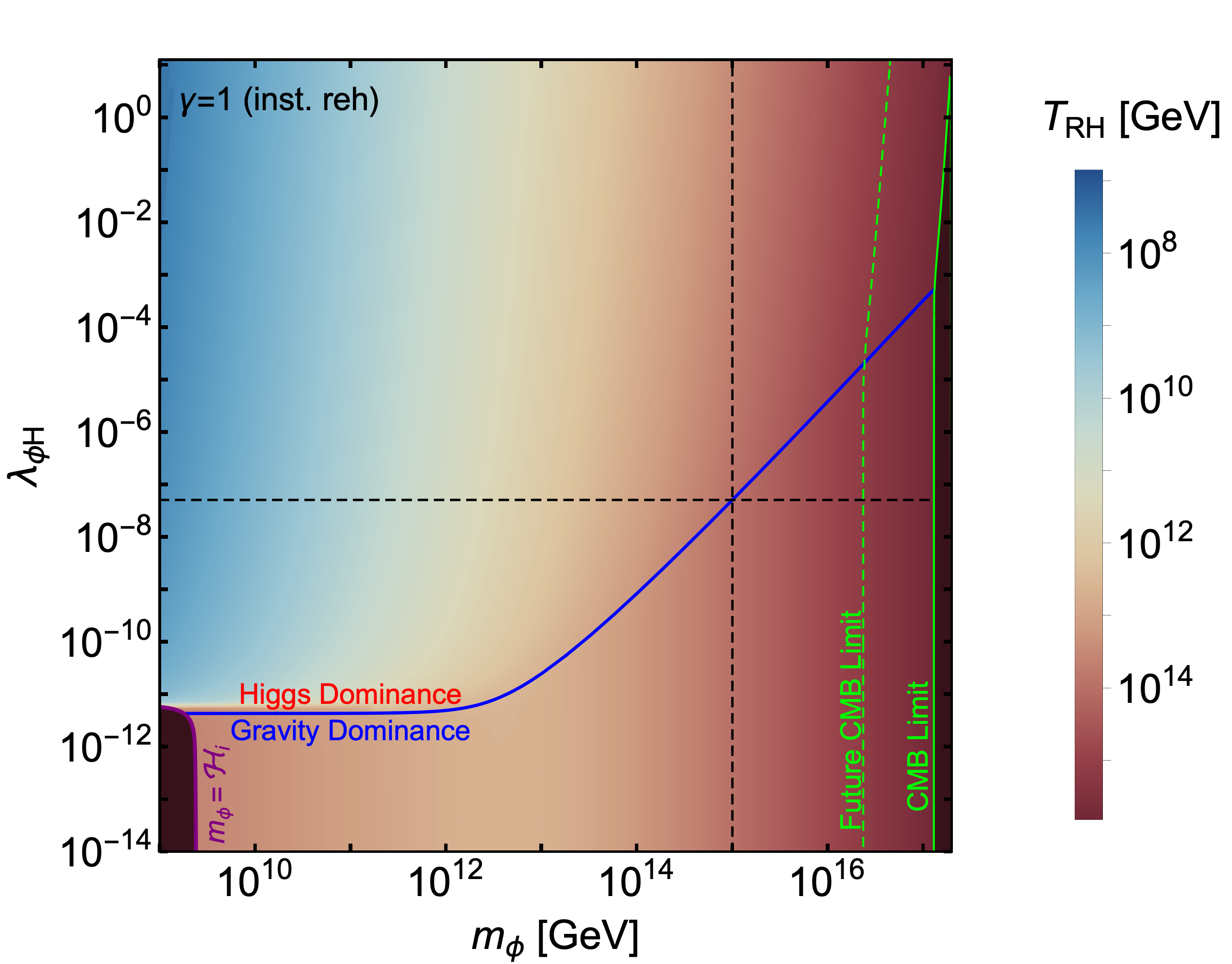}\label{fig:myT}}
\caption{\label{fig:fig2}
\textit{Left panel:}
The relation between the dark matter mass and the reheating temperature required to achieve the correct DM relic abundance, for different values of portal coupling in the instantaneous reheating scenario ($\gamma=1$). The black thick line corresponds to the pure gravity production ($\lambda_{\phi H} = 0$), while the darkest dashed blue one refers to the maximum allowed values for the coupling due to perturbativity ($\lambda_{\phi H} = 4\pi$). \rev{The shaded green region is excluded by the non-thermalisation condition~\eqref{eq:therm} when extrapolating the Higgs portal interaction above the perturbativity regime.} The dot-dashed line represents the condition $m_\phi = \Hi $, above which the parameter space is ruled out by the bound on isocurvature perturbations~\cite{Chung:2004nh,Nurmi:2015ema,Akrami:2018odb}. Moreover, the shaded grey region is excluded by the bound on tensor modes deduced by CMB data (see Eq.~\eqref{eq:limitTRH}), while the dashed line shows the sensitivity of the next-generation CMB experiments~\cite{Errard:2015cxa}.
\textit{Right panel:}
values of $m_\phi$, $\lambda_{\phi H}$ and $T_\mathrm{RH}$ achieving the correct DM relic abundance in the instantaneous reheating scenario. The blue line shows the threshold value for $\lambda_{\phi H}$ as a function of DM mass below which gravity dominates the DM production. The exclusion regions due to the bounds on tensor modes and isocurvature perturbations are delimited by green and purple lines, respectively. The green dashed line displays the future sensitivity of next-generation CMB experiments.
}
\end{center}
\end{figure}
\begin{figure}[t!]
\begin{center}
\subfigure[]{\includegraphics[width=0.48\textwidth]{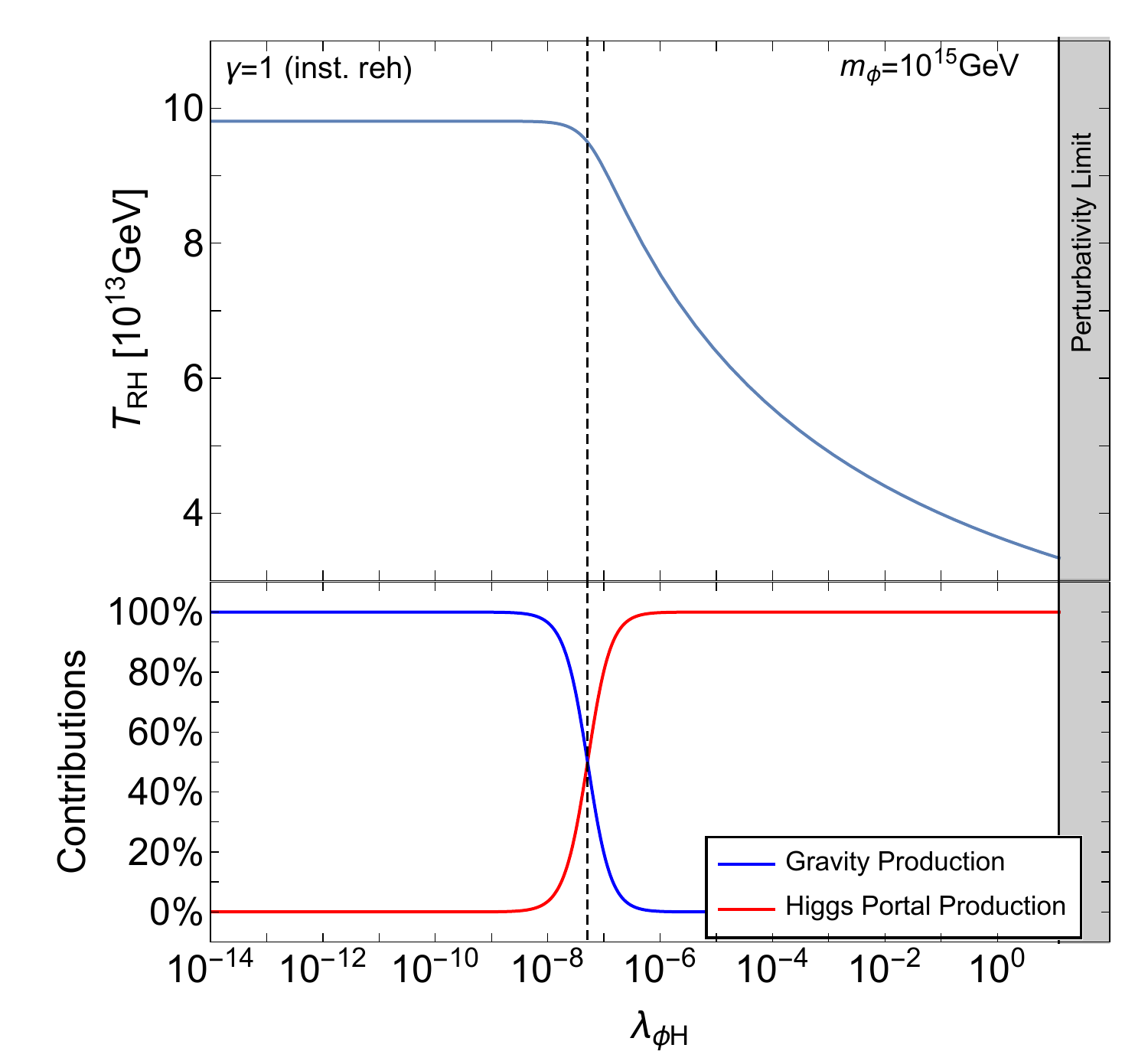}\label{fig:RelCon}}
\subfigure[]{\includegraphics[width=0.48\textwidth]{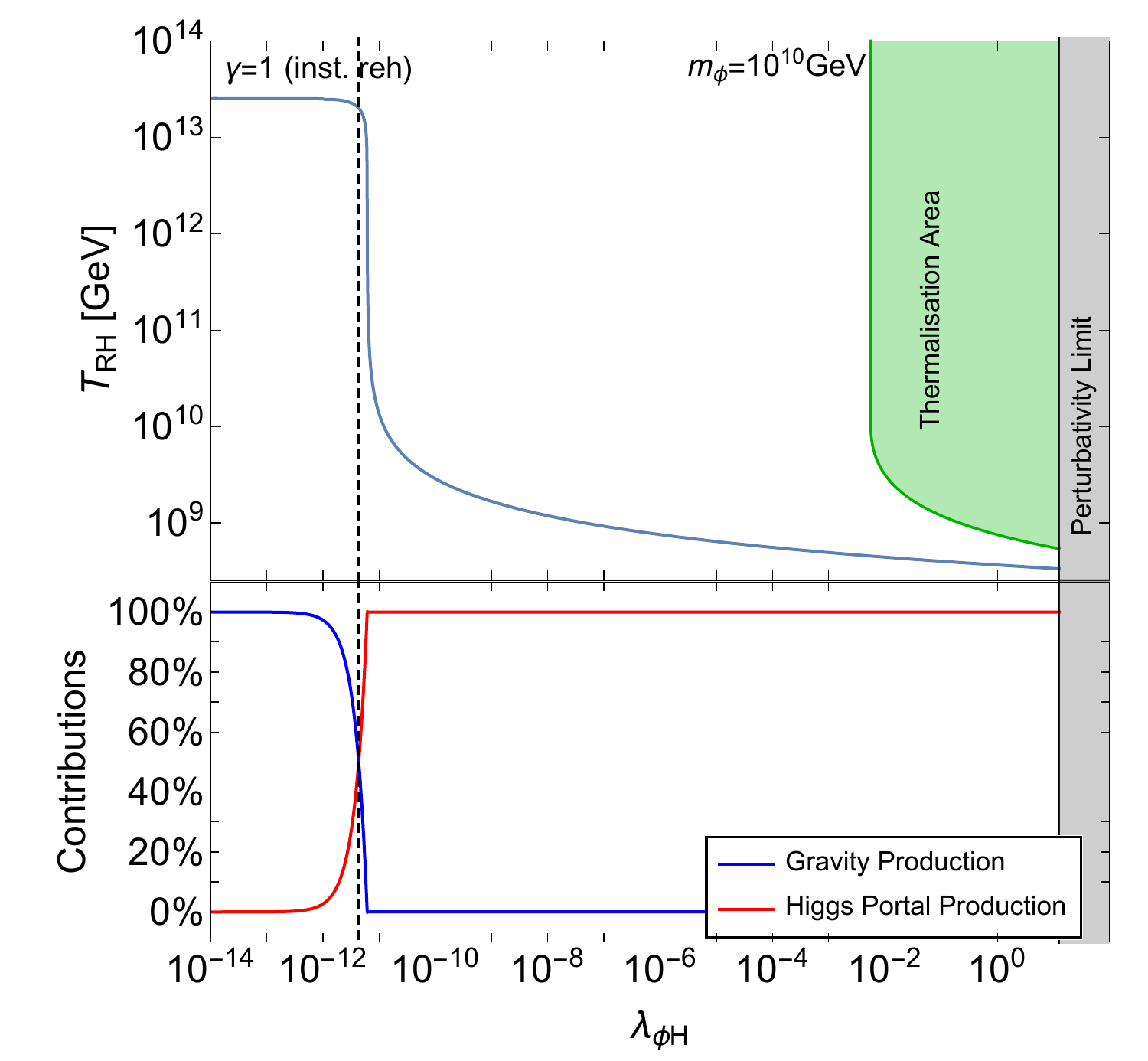}\label{fig:RelCon10}}
\caption{\label{fig:fig3}
\rev{Relation between the Higgs portal coupling and the reheating temperature achieving the correct dark matter relic abundance for the benchmark values of $m_\phi=10^{15}$~GeV (left panel) and $m_\phi=10^{10}$~GeV (right panel). The lower plots in both panels show the relative contribution of the gravity-mediated (blue line) and Higgs portal (red line) processes to the DM production.}}
\end{center}
\end{figure}

In Fig.~\ref{fig:TRHvsM1} we show the relation between the reheating temperature and the dark matter mass required to achieve the correct DM abundance for different values of the Higgs portal coupling. The black thick curve refers to the pure gravity production obtained by assuming $\lambda_{\phi H} =0$. The colored curves show how the scenario changes when turning on the Higgs portal coupling. From red to blue colors, the couplings increases so reaching the perturbativity limit (darkest dashed blue line). The lightest red line for $\lambda_{\phi H} = 5\times10^{-12}$ roughly captures the threshold value for the Higgs portal coupling. For smaller couplings, the $m_\phi$--$T_\mathrm{RH}$ curve of pure gravity production is not significantly affected, so meaning a dominance of gravity-mediated processes in DM production. In the plot, the shaded grey area is excluded by the CMB bound on the tensor-to-scalar ratio according to Eq.~\eqref{eq:limitTRH}, while the horizontal dashed line represents the sensitivity of future CMB experiments~\cite{Errard:2015cxa}. Moreover, the region above the dot-dashed black line is excluded by the limit on isocurvature perturbations~\cite{Chung:2004nh,Nurmi:2015ema,Akrami:2018odb}, since for $m_\phi > \mathcal{H}_i$ the scalar field will locally gain a vacuum expectation value $\phi_* \sim \sqrt{\left< \phi^2 \right>} = \mathcal{H}_*/2 \pi $ during the inflation.
\rev{In the plot, the shaded green region is excluded by the non-thermalisation condition~\eqref{eq:therm} when extrapolating the Higgs portal interaction above the perturbativity regime. The thermalisation constraint becomes stronger than the perturbativity constraint for $T_{\rm RH} \lesssim 1.4\times 10^{3}$ GeV, which is out of the region discussed here.}

When the scalar mass is small, the production through Higgs scattering can easily dominant. After the Higgs scattering process starts to dominant, the production at a temperature lower than the scalar mass ($T < m_\phi$) dominates. However, according to the freeze-in mechanism, we find a stronger dependence on the reheating temperature when it approaches the scalar mass. On the other hand, when the scalar mass is large, it becomes hard for the Higgs portal production to dominant because of the Boltzmann suppression. Very large Higgs portal couplings are therefore required to spoil the dominance of gravity-mediated processes. In this region, the reheating temperature is below the scalar mass, and reducing the reheating temperature substantially affects the efficiency of Higgs portal production. The Higgs portal coupling is very sensitive to the reheating temperature all over its allowed range, and in turn the perturbativity limit on the coupling highly constraints the reheating temperature.

The full scan on the three free parameters of the model ($m_\phi$, $\lambda_{\phi H}$, $T_\mathrm{RH}$) requiring the correct DM relic abundance is shown in Fig.~\ref{fig:myT}, where the reheating temperature is color-coded. The blue curve shows the threshold value for the Higgs portal coupling as a function of the DM mass. Such a value provides equal contributions of the gravity-mediated and Higgs portal scatterings to the DM production. This is the main result of the paper: below the blue line the DM production is driven by gravity-mediated processes according to the PIDM paradigm, while above the Higgs portal processes dominate. When the scalar mass is smaller than $10^{12}$~GeV, the two processes switch their dominance at around $\lambda^\mathrm{equality}_{\phi H} = 4.34 \times 10^{-12}$, which is very close to the analytical result reported in Eq.~\eqref{eq:coupling2}. For DM masses larger than $10^{12}$~GeV, the threshold value $\lambda_{\phi H}^\mathrm{equality}$ increases as the DM mass increases according to the analytical expression of Eq.~\eqref{eq:coupling1}. This behavior is due to the fact that the reheating temperature cannot be larger than the scalar mass due to the overproduction through gravity. \rev{The dependence on the reheating temperature is further investigated in Fig.~\ref{fig:fig3} for two benchmark cases with $m_\phi=10^{15}$~GeV and $m_\phi=10^{10}$~GeV. In both cases, one can see that the reheating temperature starts to decrease after the contributions from the two different production processes switch their dominance. For $m_\phi=10^{15}$~GeV (Fig.~\ref{fig:RelCon}) the switch appears at around $\lambda_{\phi H}^\mathrm{equality} = 5.1\times 10^{-8}$ (a contribution of 50\% to the DM abundance from both the two different processes), not far from the analytical estimate of $\lambda_{\phi H}^\mathrm{equality} = 2.1 \times 10^{-8}$ obtained from Eq.~\eqref{eq:coupling1}. For $m_\phi=10^{10}$~GeV (Fig.~\ref{fig:RelCon10}), we have $\lambda_{\phi H}^\mathrm{equality} = 4.34 \times 10^{-12}$ in good agreement with the analytical estimate reported in Eq.~\eqref{eq:coupling2}. Moreover, as highlighted in Fig.~\ref{fig:RelCon10}, the Higgs portal coupling required to achieve the correct dark matter relic abundance starts to vary only when $m_\phi \gtrsim T_\mathrm{RH}$ in order to balance the Boltzmann suppression. It is worth noticing that, even if large values of the Higgs portal coupling are required for $m_\phi \gg T_\mathrm{RH}$, the non-thermalisation condition reported in Eq.~\eqref{eq:therm} is satisfied and consequently the freeze-in regime is preserved as long as $T_{\rm RH} \gtrsim 1.4\times 10^{3}$ GeV.}

\begin{figure}[t!]
\begin{center}
\subfigure[]{\includegraphics[width=0.435\textwidth]{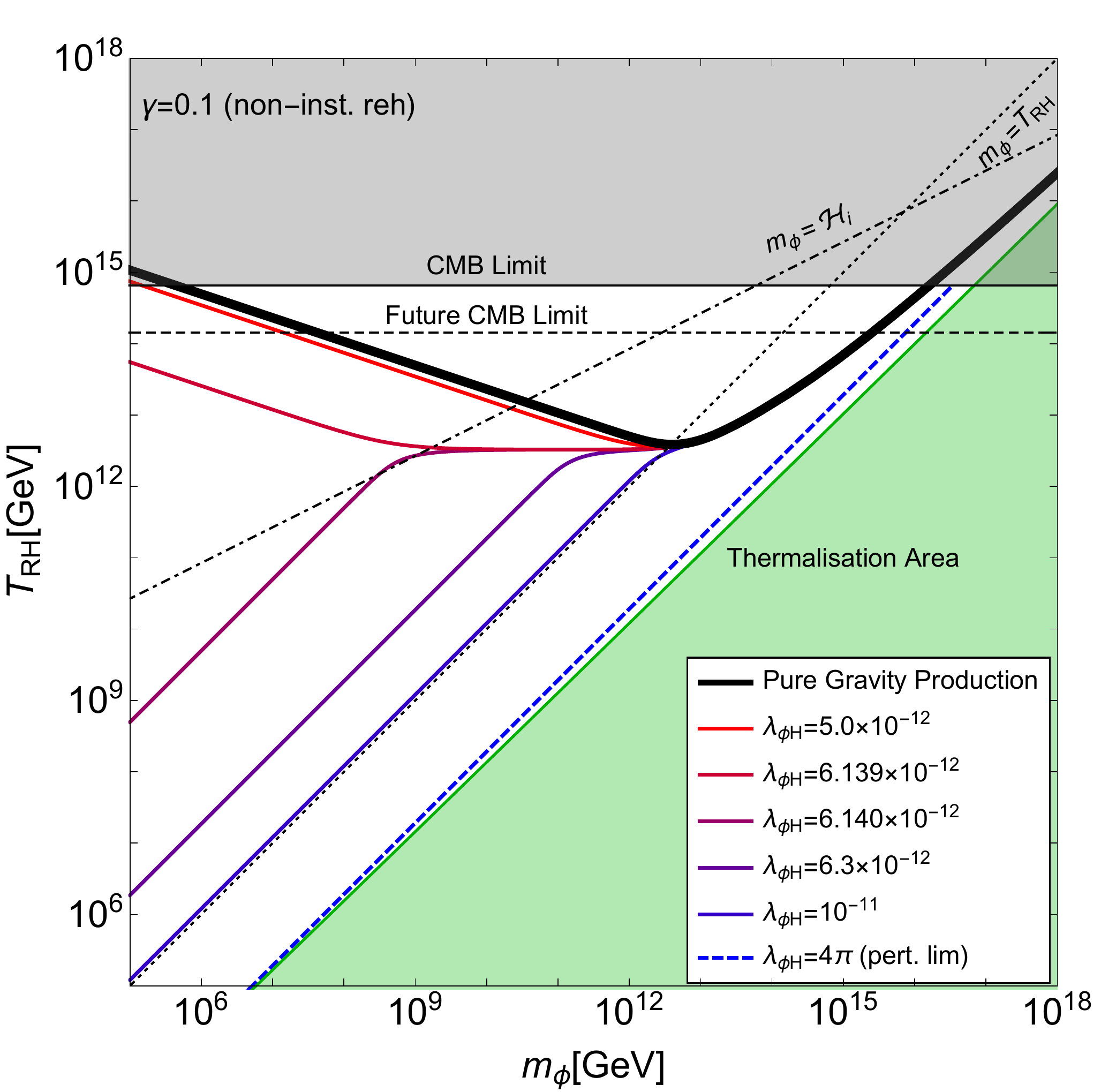}}
\subfigure[]{\includegraphics[width=0.55\textwidth]{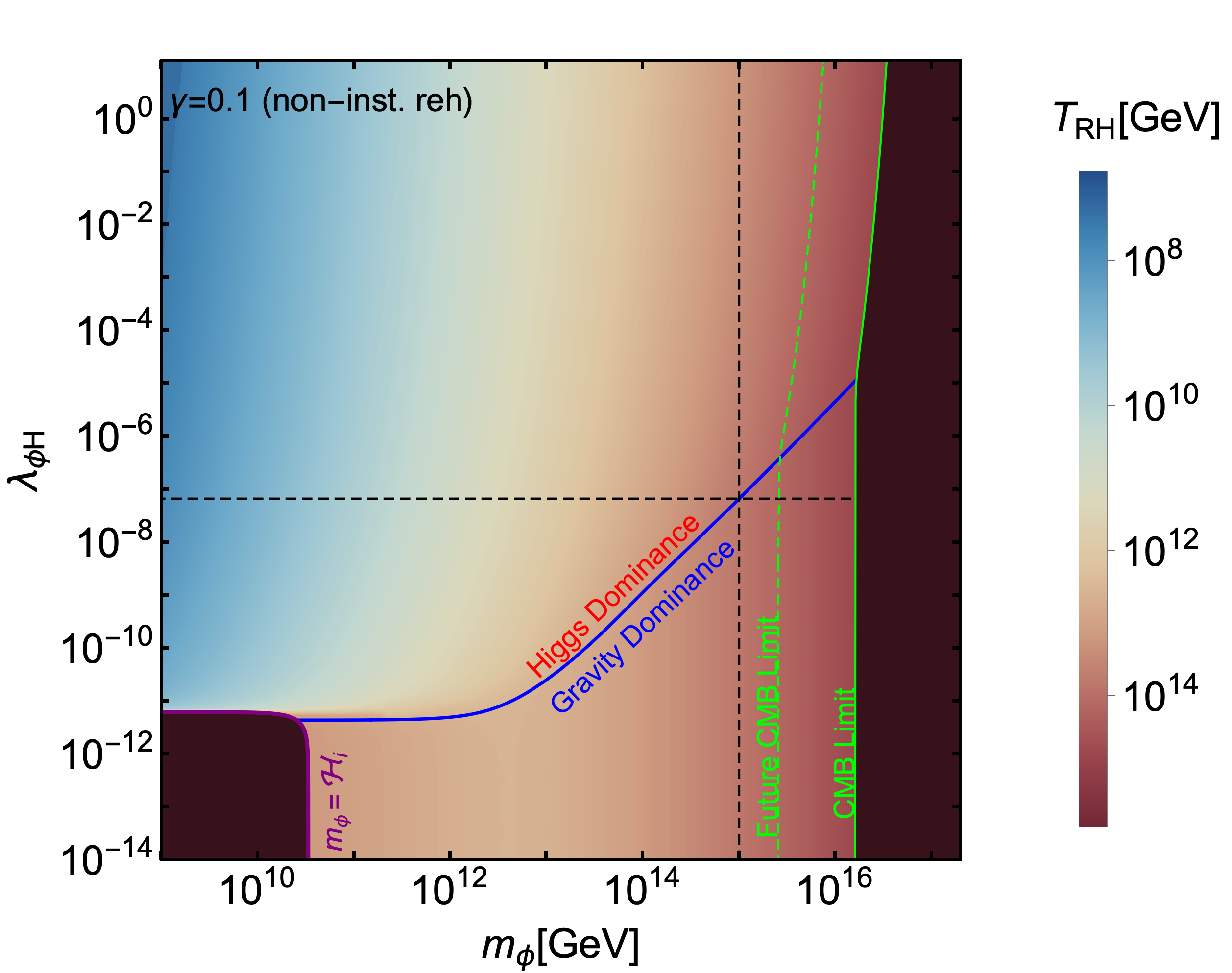}}
\caption{\label{fig:fig4}
Same as Fig.~\ref{fig:fig2} for non-instantaneous reheating scenario with $\gamma=0.1$.}
\end{center}
\end{figure}
\begin{figure}[t!]
\begin{center}
\subfigure[]{\includegraphics[width=0.435\textwidth]{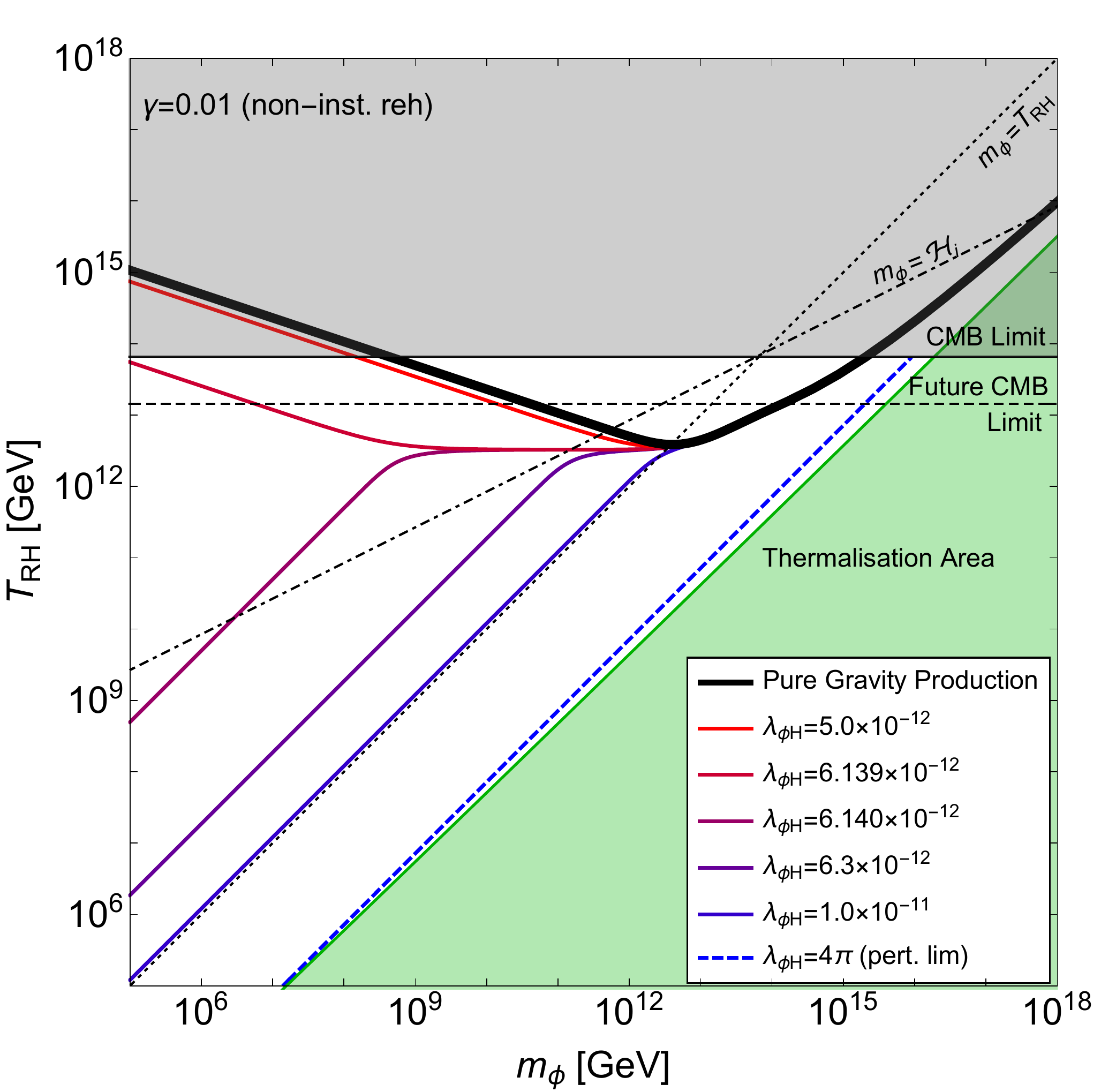}}
\subfigure[]{\includegraphics[width=0.55\textwidth]{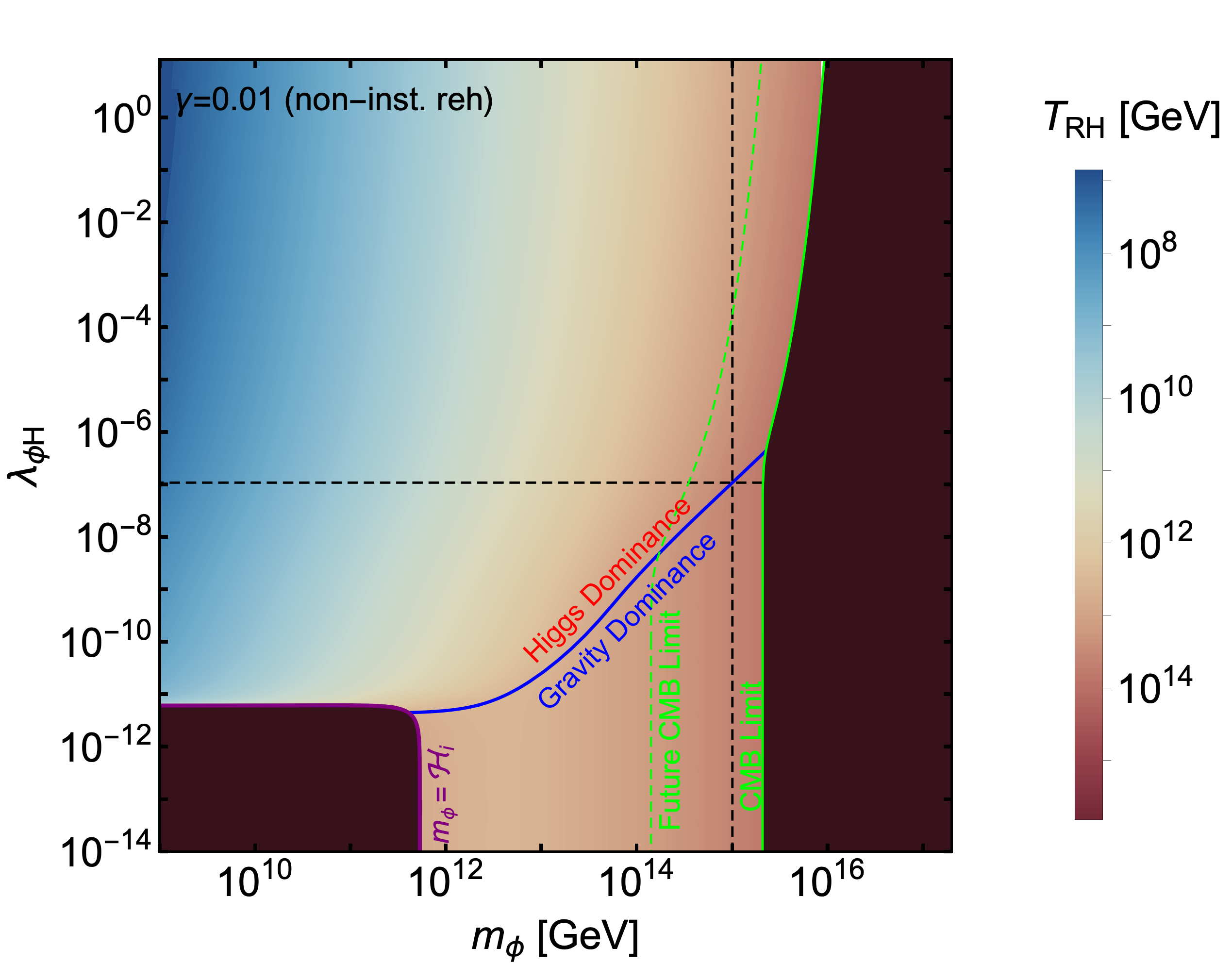}}
\caption{\label{fig:fig5}
Same as Fig.~\ref{fig:fig2} for non-instantaneous reheating scenario with $\gamma=0.01$.}
\end{center}
\end{figure}

Figures~\ref{fig:fig4} and~\ref{fig:fig5} show the non-instantaneous versions of Fig.~\ref{fig:fig2} for $\gamma=0.1$ and $0.01$, respectively. As can be seen in the plots, the relations among the free parameters of the model are very similar to the ones occurring in the instantaneous case. As a general behavior, the smaller the quantity $\gamma$, the lower the reheating temperature required to account for the correct DM abundance. Moreover, the limits from CMB data become stronger for lower values of $\gamma$, reducing the allowed region for pure gravity production. \rev{The thermalisation constraint becomes stronger than the perturbativity constraint for $T_{\rm RH} \lesssim 3.8\times 10^{3}$ GeV and $T_{\rm RH} \lesssim 8.2\times 10^{3}$ GeV for $\gamma=0.1$ and $\gamma=0.01$ respectively, which are still beyond the region discussed.} Furthermore, as shown by the blue curves in the right panels of Fig.s~\ref{fig:fig4} and~\ref{fig:fig5}, we find that the switch of the dominant contribution from the two different processes is not substantially affected by $\gamma$, even if the production during perturbative reheating is dominant for large DM mass. On the other hand, for small DM mass, this is guaranteed by that fact that DM particles are mainly produced after reheating, especially in case of the Higgs portal production. However, the approximated analytical relation in Eq.~\eqref{eq:rcontri} does not hold as good as for the instantaneous reheating scenario. Although the required reheating temperature decreases as $\gamma$ becomes smaller, the maximum temperature during reheating increases, which lowers the accuracy of Eq.~\eqref{eq:rcontri}. For instance, in case of $m_\phi=10^{15}$~GeV, the Higgs portal coupling at equality is found to be $\lambda_{\phi H}^\mathrm{equality} = 6.6\times 10^{-8}$ and $\lambda_{\phi H}^\mathrm{equality} = 1.1\times 10^{-7}$ for $\gamma=0.1$ and $\gamma=0.01$, respectively.

%%%%%%%%%%%%%%%%%%%%
%%%%%%%%%%%%%%%%%%%%
\section{Conclusions \label{sec:con}}

In the fascinating scenario known as Planckian Interacting Dark Matter (PIDM), it is generally assumed that dark matter particles have no direct coupling with the Standard Model and are produced after inflation through gravity-mediated processes. From a model building perspective, however, it is challenging to suppress the Higgs portal coupling between the Higgs and a scalar dark matter field. In the present paper, we have therefore investigated in detail the production of superheavy scalar dark matter particles in the realistic model where both the gravity-mediated and the Higgs portal processes exist. The minimal model considered has three free parameters: the reheating temperature $T_\mathrm{RH}$, the dark matter mass $m_\phi$, and the Higgs portal coupling $\lambda_{\phi H}$, with the latter controlling the production efficiency of Higgs portal processes. Moreover, we have considered the possible scenarios of instantaneous ($\gamma=1$) and non-instantaneous ($\gamma <1$) reheating. By numerically solving the Boltzmann equations in the freeze-in limit during and after reheating, we have provided the relations among the model parameters required to account for the correct DM relic abundance.

Most importantly, we have highlighted the regions of the parameter space where one of the two production processes (gravity and/or Higgs scatterings) dominates the production of dark matter particles. In particular, we have estimated a threshold value for the Higgs portal coupling below which the corresponding processes are sub-dominant, so preserving the assumptions of the PIDM scenario. We have provided analytical expressions for such a threshold value by requiring the equality of the contributions to the dark matter yield from the two different processes. The accuracy of the analytical results has been numerically tested. In the case of instantaneous reheating, for the benchmark dark matter mass of $10^{15}$~GeV, we have found that the Higgs portal coupling has to be smaller than $5.1 \times 10^{-8}$. This upper bound is relaxed by a factor of 2 in case of non-instantaneous reheating. On the other hand, it is worth noticing that the regions where the Higgs portal processes dominate are less constrained by current and future CMB limits. In particular, superheavy dark matter particles with a mass larger than $10^{15}$~GeV are still allowed in the case of large values for the Higgs portal coupling.

%%%%%%%%%%%%%%%%
\section*{Acknowledgments}
BF acknowledges the Chinese Scholarship Council (CSC) Grant No.\ 201809210011 under agreements [2018]3101 and [2019]536. SFK acknowledges the STFC Consolidated Grant ST/L000296/1 and the European Union's Horizon 2020 Research and Innovation programme under Marie Sk\l {}odowska-Curie grant agreements Elusives ITN No.\ 674896 and InvisiblesPlus RISE No.\ 690575.

\bibliography{DarkMatter}

\end{document}